\documentclass{spieman}
\usepackage{graphicx} 
\usepackage{amsmath,amsfonts,amssymb}
\usepackage[square,numbers]{natbib}
\title{Comments concerning the paper ``On the calibration of ultra-high energy EASs at the Yakutsk array and Telescope Array’’ by A.V.~Glushkov \textit{et al.} }
\author[1]{John N. Matthews}
\author[2]{Yoshiki Tsunesada}
\author{: on behalf of the Telescope Array Collaboration}
\affil[1]{Department of Physics and Astronomy and the High Energy Astrophysics Institute, University of Utah, Salt Lake City, Utah 84112, USA}
\affil[2]{Graduate School of Science and Nambu-Yoichiro Institute for Theoretical and Experimental Physics, Osaka Metropolitan University, Osaka 558-8585, Japan}

\begin{document}

\maketitle

We recently reviewed a paper authored by the Yakutsk Group focusing on the energy scales of the Yakutsk and Telescope Array (TA) experiments. The paper is intended for publication in the Journal \textit{Physics of Atomic Nuclei} \cite{Yak1}. In their study, the authors developed a customized detector response simulator that incorporates various physical processes including ionization, bremsstrahlung, pair production, and Compton scattering. They applied this simulator to both Yakutsk and Telescope Array surface detectors and reached a conclusion that the TA energy scale might be wrong due to an incorrect definition of the ``response unit''.

The authors refer to the TA's ``energy deposit formula'', expressed as 
\begin{align}
    \mathrm{VEM} &= 2.05 \times \sec \theta \; \mathrm{MeV}. \label{eq6}
\end{align}
This equation (numbered (6) in the Yakutsk paper) was originally given in Dmitri Ivanov's Ph.D.~thesis at Rutgers University \cite{Dmitri} (cited as reference [13] in their paper).  The authors also introduced equation (10) in the paper which scales the equation above by two factors, 1.2 and 1.036:
\begin{align}
    \mathrm{VEM} &= 2.05 \times 1.2 \times 1.036 \times \sec \theta \; \mathrm{MeV}. \label{eq10}
\end{align}
They attributed the two factors to the thickness (in cm) and the density (in $\mathrm{g/cm^3}$) of the scintillator material used in the TA surface detector \cite{TASD}. 

The authors conducted a comparison of simulation results employing their own detector response calculator on shower particles generated utilizing the hadronic interaction model QGSJET-II-04. They found agreement between their result and the calculation by TA for vertical showers, but not for inclined showers. They attributed this discrepancy to their conclusion \textit{``incorrectly chosen unit of VEM = 2.05 MeV in \cite{Dmitri}, which does not reflect the real physical processes occurring in the scintillation detector during registration of EAS''}. However, we note that this conclusion is incorrect.

The energy deposit formula \eqref{eq6} given in Ivanov \cite{Dmitri}, equation (6) of the Glushkov {\it et al.} paper, was derived from the TA’s detector Monte-Carlo simulation using the
widely-used GEANT4 package, commonly employed in high-energy physics and cosmic ray studies. The TA detector simulator incorporates all the details of the TA surface detector including the structure, housing, and the scintillator material as well as its dimensions. We computed the energy deposit by charged particles in an air shower impacting various positions within the $3 \, \mathrm{m^2}$ area scintillator. The value of 2.05 MeV represents the most probable energy deposit by an individual charged particle traversing vertically through the 1.2 cm thick scintillator. We also confirmed that this value scales as $\sec \theta$ with the incident angle of the particle.

It is important to note that the value 2.05 MeV is expressed in terms of the energy measured in MeV, already accounting for the thickness and density of the TA scintillator. It does not represent the energy deposit per unit length. Therefore, scaling this value by the thickness or the density of the scintillator is not meaningful. The assertion made by Glushkov {\it et al.} stems from a misinterpretation and an incorrect application of the TA energy deposit formula. Most importantly, we emphasize that the Telescope Array detector simulations have been validated by comparisons with the actual observed data: we found excellent agreement between the TA’s Monte-Carlo prediction of the detector outputs and the real data, including the distribution of signal amplitudes \cite{TASDsp}.

\end{document}